\documentclass[11pt,a4paper]{article}

\def\AmSTeX{\leavevmode\hbox{$\mathcal A\kern-.2em\lower.376ex%
        \hbox{$\mathcal M$}\kern-.2em\mathcal S$-\TeX}}
\usepackage{graphicx}
\usepackage{amsmath}        
\usepackage{amssymb}
\usepackage{mathrsfs}       
\usepackage{bm}             
\newif\ifpdf \pdftrue
\ifx\pdfoutput\undefined \pdffalse \fi \ifx\pdfoutput\relax
\pdffalse \fi
\ifx\texonly\undefined\let\texonly\relax\fi
\ifx\endtexonly\undefined\let\endtexonly\relax\fi \texonly
  \let\htmlonly\iffalse
  \let\endhtmlonly\fi
\endtexonly

\htmlonly
  
\endhtmlonly
\texonly
  \usepackage{makeidx}

  \usepackage{multicol}
  

\endtexonly
\title{}
\author{\thanks{}}
\date{}
\begin{document}

\title{Total Hadronic Cross Sections and $\pi^\mp \pi^+$ Scattering}

\author{Francis Halzen$^a$\footnote{francis.halzen@icecube.wisc.edu},~~
Keiji Igi $^b$\footnote{igi@phys.s.u-tokyo.ac.jp},~~
Muneyuki Ishida $^c$\footnote{mishida@wisc.edu},~~
C. S. Kim $^d$\footnote{cskim@yonsei.ac.kr,~ Corresponding Author}\\
{\it \small  $^a$ Department of Physics, University of Wisconsin, Madison, WI 53705, USA} \\
{\it \small  $^b$ Mathematical Physics Laboratory, RIKEN Nishina Ctr., Wako, Saitama 351-0198, Japan} \\
{\it \small  $^c$ Department of Physics, Meisei University, Hino, Tokyo 191-8506, Japan} \\
{\it \small  $^d$ Department of Physics $\&$ IPAP, Yonsei
University, Seoul 120-749, Korea}}

\maketitle

\baselineskip=20pt

\begin{abstract}

\noindent Recent measurements of the inelastic and total proton-proton cross section at the LHC,
and at cosmic ray energies by the Auger experiment, have quantitatively confirmed fits to lower energy data
 constrained by the assumption that the proton is asymptotically a black disk of gluons.
We show that data on $\bar p(p)p,\pi^\mp p$, and $K^\mp p$ forward scattering support the related expectation
 that the asymptotic behavior of all cross sections is flavor independent.
By using the most recent measurements from ATLAS, CMS, TOTEM and Auger, we predict
$\sigma^{pp}_{\rm tot} (\sqrt s=8~{\rm TeV})=100.6 \pm 2.9$ mb and
$\sigma^{pp}_{\rm tot} (\sqrt s=14~{\rm TeV})=110.8 \pm 3.5$ mb,
as well as refine the total cross section
$\sigma^{pp}_{\rm tot} (\sqrt s=57~{\rm TeV})=139.6 \pm 5.4$ mb.
Our analysis also predicts the total $\pi^\mp \pi^+$ cross sections
as a function of $\sqrt s$.
\end{abstract}

\hspace*{0.5cm}PACS : 13.85.Lg\ \ 13.75.Cs\ \ 14.20.Dh

\vspace*{-12cm}

\hspace*{12cm}RIKEN-MP-27

\newpage

\section{Introduction}

Recent high energy measurements of the inelastic proton-proton cross section, made possible by the Large Hadron Collider (LHC) and a new generation of cosmic ray experiments, have convincingly confirmed \cite{Block:2011vz} indications \cite{P1,COMPETE,BH,IgiIshida} in lower energy data that the total cross section $\sigma_{\rm tot}$ behaves asymptotically as the squared log of the center-of-mass energy $\sqrt s$, reminiscent of the energy dependence of Froissart's unitarity bound \cite{P2}. This energy dependence is now solidly anchored to all $pp$ and $\bar p p$ total and inelastic cross section measurements, from threshold data averaged by finite energy sum rules, to the result at 57\,TeV center-of-mass energy of the Auger cosmic ray array \cite{IgiIshida}.

The energy dependence is suggestive of that predicted by an asymptotic black disk. Although the data itself does not cover asymptotic energies, from an extrapolation of the fits constrained by analyticity, the features of a black disk emerge, with a purely imaginary amplitude and a ratio of $\sigma_{inel}/\sigma_{tot}$ consistent with 0.5 within errors \cite{Block:2011vz}. Additional, and independent, confirmation has been provided by LHC measurements of the shrinkage of the elastic scattering cross section \cite{Schegelsky:2011aa}. From a parton point of view, the picture that emerges asymptotically is that of a proton composed of an increasing number of soft gluon constituents, each carrying a decreasing fraction of the proton energy. The asymptotic cross section, clearly emerging from available data is given by
\begin{eqnarray}
\sigma_{tot} &=& {4 \pi \over M^2}\,\, ln^2 \frac{s}{s_0},
\end{eqnarray}
where M, historically identified with the mass of the pion, is now associated with the particles populating the Pomeron trajectory, i.e. glueballs.

If one ascribes the origin of the asymptotic $ln^2 s$ term in $\bar p p$ and $pp$ scattering to gluons only, then it is universal and its energy dependence as well as its normalization is  the same for $\pi \pi$, $\pi p$, $Kp$, and $\gamma p$ interactions via vector meson dominance. In other words, the role of quarks, and therefore the quantum numbers of hadrons, becomes negligible. Although there is still no rigorous derivation, the straightforward interpretation of the present data is that, asymptotically, particles of all flavors evolve into a universal black disk of gluons. The COMPETE Collaboration already proposed that this asymptotic behavior $\sigma_{\rm tot}\simeq B~{\rm log}^2(s/s_0)$ applies to all hadron total cross sections, with a universal value of the coefficient $B$ \cite{COMPETE,PDG}.

In order to empirically test this universality, the $\bar p(p)p,\pi^\mp p$, and $K^\mp p$
forward scattering amplitudes are analyzed, and the values of $B$, denoted respectively as
$B_{pp}, B_{\pi p},$ and $B_{Kp}$, were estimated independently \cite{II}.
The analysis was refined \cite{IB} for $B_{Kp}$. The resulting values are consistent with the
universality, $B_{pp}\simeq B_{\pi p}\simeq B_{Kp}$, and thus,
the universality of $B$ is suggested.
Recently strong indications for a universal and Froissart-like hadron-hadron total cross
section at high energy are also obtained in the lattice
QCD simulations \cite{lattice}.
In this work
we first update the analysis of the $\bar p(p)p,\pi^\mp p$, and $K^\mp p$ data by including newly measured LHC results as well as very high energy measurements based on cosmic ray data.
We also fit the $\bar p(p)n$ data at the same time.
Subsequently, assuming the universality of $B$, we calculate the $\pi^\mp \pi^+$ total cross section
$\sigma_{\rm tot}^{\pi^\mp \pi^+}(s)$ at all energies. Similar analyses are also done
in Refs. \cite{PY,Aichili,Leutwyler} by using different methods.

Although challenging, the data on $\pi^\mp \pi^+$ collisions could be extended to higher energies exploiting high intensity proton beam accelerator beams planned worldwide,  such as Project X \cite{Project-X} of FNAL and J-PARC in Japan \cite{J-Parc}. At a later stage these may develop into muon colliders. As an example, Project X, a high intensity proton source proposed at Fermilab,
would deliver proton beams at energies ranging from 2.5 to 120 GeV \cite{Project-X} and secondary pion beams with $E(\pi) \approx 2 - 15$ GeV. A muon collider with Project-X-intensity pion beams would represent a $\pi^+ \pi^-$ collider with $\sqrt s= 1$ TeV and a luminosity of $10^{22} cm^{-2}/sec$ \cite{SGeer}, not quite sufficient, even for measuring the large cross sections discussed here. Some manipulation of the secondary beams would be required. On the other hand, direct measurements of $\sigma_{\rm tot}^{\pi\pi}$ in wide range of pion beam energy would be made possible.
In the absence of such measurements we will extend our calculations of $\sigma_{\rm tot}^{\pi^\mp\pi^+}(s)$ into the intermediate energy
region using Regge theory. This will allow us to compare our predictions with indirect information \cite{Ro,Bi,Co,PY,PY2,Za,Ha,Ab} extracted from processes
such as $\pi^- p\rightarrow \pi^-\pi^+ n,\ \pi^-\pi^-\Delta^{++}$,
assuming one-pion-exchange dominance.

\section{Update of the fits to $\sigma_{\rm total}$}

\subsection{Analysis of Forward $\bar p(p)p,\pi^\mp p,K^\mp p,\bar p(p)n$ Amplitudes}

The energy (momentum) of the beam in the laboratory system is denoted by $\nu (k)$. It is related
to the center of mass energy $\sqrt s$ by
\begin{eqnarray}
s &=& 2 M \nu + M^2 + m^2, \ \ \ \nu=\sqrt{k^2+m^2} \ \ ,
\label{eqP1}
\end{eqnarray}
where $m=M,\mu,m_K$ for $pp,\pi p,Kp$ scattering, and $M$, $\mu$, and $m_K$ are proton, pion and kaon masses;
respectively.
$s\simeq 2M\nu$ in high-energies. For $\bar p(p)n$, $M$ is replaced by neutron mass $M_n$
and $m=M$.

The crossing-even forward scattering amplitude, $F_{ab}^{(+)}(\nu )$, is given by
the sum of Pomeron and Reggeon (including $P^\prime$ trajectory) exchange terms, while
the crossing-odd $F_{ab}^{(-)}(\nu )$ is given by
a single contribution from Reggeon (corresponding to vector-meson trajectories) exchange contributions.
Here the subscripts $ab$ and $\bar ab$ represent $ab=pp,\pi^+p,K^+p,pn$ and
$\bar ab=\bar pp,\pi^-p,K^-p,\bar pn$, respectively.
We consider the exchange degenerate $f_2(1270)$-, $a_2(1320)$-trajectories for the crossing-even Reggeon (tensor-meson) term and the $\rho$-, $\omega$-trajectories for the vector-meson term.
Their imaginary parts are given explicitly by
\begin{eqnarray}
Im~F_{ab}^{(+)}(\nu ) &=&  \frac{\nu}{m^2}\left( c_2^{ab} {\rm log}^2\frac{\nu}{m}
   + c_1^{ab} {\rm log}\frac{\nu}{m} + c_0^{ab}  \right)
   + \frac{\beta_{T}^{ab}}{m}\left( \frac{\nu}{m} \right)^{\alpha_{T}(0)}~,
\label{eqP2}\\
Im~F_{ab}^{(-)}(\nu ) &=&  \frac{\beta_V^{ab}}{m}\left( \frac{\nu}{m} \right)^{\alpha_{V}(0)}\ ,
\label{eqP3}
\end{eqnarray}
where $c_0^{ab}$, $\beta_{T}^{ab}$, and $\beta_V^{ab}$ are unknown parameters
in the Pomeron-Reggeon exchange model.
The $c_2^{ab}$ and $c_1^{ab}$ are introduced consistently with the Froissart bound to describe the
increase of $\sigma_{\rm tot}$ at high energy.
The intercepts are fixed with $\alpha_{T}(0)=0.542,\ \alpha_{V}(0)= 0.455$,
which is taken to be the same as the Particle Data Group \cite{PDG}.
The amplitudes Im~$F_{ab}^{(\pm)}(\nu)$ are related to the total cross sections $\sigma_{\rm tot}^{\bar ab,ab}(s)$ by the optical theorem:
\begin{eqnarray}
\sigma^{\bar a b}_{\rm tot}(s) &=& \sigma_{ab}^{(+)}(s) + \sigma_{ab}^{(-)}(s)\ ,\ \
\sigma^{ab}_{\rm tot}(s) = \sigma_{ab}^{(+)}(s) - \sigma_{ab}^{(-)}(s)~,\nonumber\\
{\rm where}&&\sigma_{ab}^{(\pm )}(s) \equiv \frac{4\pi}{k}\ Im~F_{ab}^{(\pm )}(\nu )\ \ .
\label{eqP4}
\end{eqnarray}

In our analysis, $\rho^{\bar ab,ab}(s)$, the ratios of real to imaginary parts of
forward amplitudes, are fitted simultaneously with the data on $\sigma_{\rm tot}^{\bar ab,ab}$.
Real parts of the crossing-even$/$odd amplitudes are directly obtained from crossing symmetry
$F^{(\pm )}(e^{i\pi}\nu )=\pm F^{(\pm)}(\nu)^*$ as
\begin{eqnarray}
Re~F_{ab}^{(+)}(\nu) &=& \frac{\pi\nu}{2m^2}\left( c_1^{ab} + 2 c_2^{ab} {\rm log}\frac{\nu}{m} \right)
-\frac{\beta_{T}^{ab}}{m} \left( \frac{\nu}{m} \right)^{\alpha_{T}(0)}
{\rm cot}\frac{\pi\alpha_{T}(0)}{2}+F_{ab}^{(+)}(0)\ ,\label{eqP5}\\
Re~F^{(+)}(\nu) &=& \frac{\beta_{V}^{ab}}{m}\left(\frac{\nu}{m}\right)^{\alpha_{V}(0)}
{\rm tan}\frac{\pi\alpha_{V}(0)}{2}\ .
\label{eqP6}
\end{eqnarray}
We introduce $F_{ab}^{(+)}(0)$ as a subtraction constant in the
dispersion relation \cite{Block}. The $\rho^{\bar ab,ab}(s)$ are given by
\begin{eqnarray}
\rho^{\bar ab,ab}(s) &=& Re~F^{\bar ab,ab}(\nu)/Im~F^{\bar ab,ab}(\nu),\ \
F^{\bar ab,ab}(\nu)=F_{ab}^{(+)}(\nu) \pm F_{ab}^{(-)}(\nu )\ \ .
\label{eqP7}
\end{eqnarray}

\subsection{Constrained Analysis with Universal Rise of $\sigma_{\rm tot}$ and Duality}

The contributions of the tensor term in Eq.~(\ref{eqP2}) and the vector term of Eq.~(\ref{eqP3})
to the $\sigma_{\rm tot}^{\bar ab,ab}(s)$ are negligible  in the high-energy limit
$\sqrt s\rightarrow\infty$, where they are well approximated by the $c_{2,1,0}^{ab}$ terms:
\begin{eqnarray}
\sigma_{\rm tot}^{\bar ab}(s)\simeq \sigma_{\rm tot}^{ab}(s)
   &\simeq& B_{ab}{\rm log}^2\frac{s}{s_0^{ab}} + Z_{ab}~,
\label{eqP8}\\
{\rm where}&&
B_{ab} = \frac{4\pi}{m^2}c_2^{ab},\ \ \ Z_{ab}=\frac{4\pi}{m^2}\left( c_0^{ab}-\frac{c_1^{ab\ 2}}{4c_2^{ab}} \right) ~,
\label{eqP9}\\
&& s_0^{ab} = 2M\nu_0^{ab}+M^2+m^2,\ \nu_0^{ab}=m\ e^{-\frac{c_1^{ab}}{2c_2^{ab}}}~,
\label{eqP10}
\end{eqnarray}
where $s_0^{ab}$ is a scale for the collision energy squared. By neglecting the small tensor-term contribution, $\sigma^{\bar ab,ab}_{\rm tot}$ develops a minimum $Z_{ab}$. $B_{ab}$ controls the increase of
$\sigma_{\rm tot}^{\bar ab,ab}(s)$ at high energy.
In practise, tensor and vector contributions
are negligible for $\sqrt s \stackrel{>}{\scriptscriptstyle \sim} 50$GeV where $\sigma_{\rm tot}^{\bar ab}$ and $\sigma_{\rm tot}^{ab}$
are described by Eq.~(\ref{eqP8}).

Two independent analyses \cite{II,IB} of forward $\bar p(p)p,\pi^\mp p,K^\mp p$ scattering
using finite-energy sum rules (FESR) as constraints, demonstrated that the
universality relation $B_{pp}=B_{\pi p}=B_{Kp}$ is valid to within one standard deviation.
In the present analysis, we include the $\bar p(p)n$ data and assume
this universality from the beginning.
\begin{eqnarray}
B_{pp}=B_{\pi p}=B_{Kp}=B_{pn} &\equiv& B\
\label{eqP11}
\end{eqnarray}
It leads to
constraints among $c_2^{pp}$, $c_2^{\pi p}$, $c_2^{Kp}$, and $c_2^{pn}$ from Eq.~(\ref{eqP9}).


Other powerful constraints are obtained from FESR \cite{IgiIshida}
for crossing-even amplitudes,
\begin{eqnarray}
\frac{2}{\pi}\int_{N_1}^{N_2}\frac{\nu}{k^2}{\rm Im}F_{ab}^{(+)}(\nu)d\nu &=&
\frac{1}{2\pi^2}\int_{\overline{N_1}}^{\overline{N_2}}\sigma_{ab}^{(+)}(k)dk =
\frac{1}{2\pi^2}\int_{\overline{N_1}}^{\overline{N_2}}
  (\sigma_{\rm tot}^{\bar ab}(k)+\sigma_{\rm tot}^{ab}(k))/2\ dk ~,
\label{eqP12}
\end{eqnarray}
where
$\overline{N_{1,2}}=\sqrt{N_{1,2}{}^2-m^2}$. The integration limit $N_2$ is taken in the asymptotically high-energy region, while $N_1$ is in the resonance-energy region.
The left hand side of Eq.~(\ref{eqP12}) is calculated analytically from the asymptotic formula of
Im~$F^{(+)}_{ab}$ given by Eq.~(\ref{eqP2}), while the right hand side is estimated from low-energy experimental data.
Equation (\ref{eqP12}) imposes duality on the analysis. It allows us to constrain the high-energy asymptotic behavior with the very precise low energy data, through averaging of the resonances.

Following Ref. \cite{II}, we take $\overline{N_1}=0.818,~5,~5$ GeV for $ab=\pi p,~pp,~Kp$,
while $\overline{N_2}$ is commonly taken as $\overline{N_2}=20$ GeV. The FESR (\ref{eqP12}) yields \cite{II} the constraints,
\begin{eqnarray}
(\pi p) && 102.2\beta_{T}^{\pi p}+627.3c_0^{\pi p}+2572c_1^{\pi p}+10891c_2^{\pi p}=66.96\pm 0.04,
\label{eqP14}\\
(K p) && 9.353\beta_{T}^{K p}+39.23c_0^{K p}+124.1c_1^{K p}+398.5c_2^{K p}=38.62\pm 0.07,
\label{eqP15}\\
(p p) && 3.481\beta_{T}^{p p}+10.89c_0^{p p}+27.50c_1^{p p}+71.00c_2^{p p}=90.38\pm 0.20.
\label{eqP16}
\end{eqnarray}
The integrals of the experimental cross sections in the right hand side are estimated very accurately from low-energy data with errors less than 1\%,
and these equations can be regarded as exact constraints among parameters.

\subsection{Updated Analysis Including LHC and Very High Energy Cosmic-Ray Data}

In order to determine the value of $B$ more precisely,
we now include three recent measurements, ATLAS, CMS and Auger, covering the very high-energy region in our fit:

\begin{itemize}
\item ATLAS reported \cite{ATLAS} a $pp$ inelastic cross section $\sigma_{\rm inel}^{pp}$ at 7 TeV
of $69.4\pm 2.4({\rm exp.})\pm 6.9({\rm extr.})$ mb where exp./extr. refers to errors from experimental/extrapolation
uncertainties. By using the ratio $\sigma_{\rm tot}/\sigma_{\rm inel}$ at 7 TeV of 1.38, obtained from the eikonal model \cite{BHinel}, $\sigma_{\rm tot}^{pp}$ is predicted to be
$\sigma_{\rm tot}^{pp}({\rm 7~TeV})=96.0\pm 3.3\pm 9.5$ mb.
Recently, this measurement was confirmed by the CMS collaboration \cite{CMS} reporting
$\sigma_{\rm inel}=68.0\pm 2.0({\rm syst.})\pm 2.4({\rm lum.})\pm 4({\rm extr.})$ mb, (where lum. refers to the error associated with the luminosity) giving $\sigma_{\rm tot}^{pp}=94.0\pm 2.8\pm 3.3\pm 5.5$ mb at the same energy.
We include these data omitting extrapolation errors.

\item The Auger \cite{Auger} collaboration measured $\sigma_{\rm inel}^{pp}$ at 57 TeV to be
$90\pm 7({\rm stat.}) \stackrel{+8}{\scriptstyle -11}({\rm syst.})  \pm 1.5({\rm Glauber})$, where
the last contribution to the error comes from Glauber theory.
Using $\sigma_{\rm tot}/\sigma_{\rm inel}=1.45$ at 57~TeV from Ref. \cite{BHinel}, $\sigma_{\rm tot}^{pp}$ at 57~TeV is predicted to be $131\pm 10 \stackrel{+12}{\scriptstyle -16}  \pm 2$ mb.
We also include this result with statistical error only.

\item The TOTEM \cite{TOTEM} has measured a total proton-proton cross section
at $\sqrt s = 7$ TeV,
$98.3 \pm 0.2({\rm stat.}) \pm 2.8({\rm syst.})$ mb.
\end{itemize}

Experimental data of $\sigma_{\rm tot}^{\bar ap,ap}$ at $k\ge 20$~GeV and
$\rho^{\bar ap,ap}$ at $k\ge 5$~GeV for  $\bar p(p)p,\pi^\mp p,K^\mp p$ scattering
are analyzed. We also include the data of $\sigma_{\rm tot}^{\bar pn,pn}$ and
$\rho^{pn}$ at $k\ge 10$~GeV. These data are fit simultaneously
imposing on the parameters $c_{2,1,0}^{ap},\beta_{T ,V}^{ap},F^{(+)}_{ap}(0)$
the constraints (\ref{eqP11}) and (\ref{eqP14}-\ref{eqP16}).
The highest energy data for $\sigma_{\rm tot}^{\bar ap,ap}$ data reach 26.4(25.3)~GeV for $\pi^-p(\pi^+p)$,
24.1~GeV for $K^\mp p$, 1.8~TeV for $\bar pp$ (Tevatron), 57~TeV for $pp$ (Cosmic-Ray),
and 23.0(26.4)~GeV for $\bar pn(pp)$.

\begin{table}[htb]
\caption{Best-fit parameters of the fit to $\sigma_{\rm tot}$ and $\rho$-ratios of
$\pi^\mp p$, $K^\mp p$, and $\bar p(p)p$ scatterings. Constraints of the universality of $B$,
Eq.~(\ref{eqP11}) and FESR (\ref{eqP14}-\ref{eqP16}) are used. The brackets represent the most
dominant uncertainties: the statistical errors for $\beta_V$ and $F^{(+)}(0)$ and
the systematic errors, which comes from the TOTEM measurement \cite{TOTEM},
for the other parameters.
}
\begin{tabular}{c|cccc|cc}
\hline\hline
$ab$ & $B$(mb) & $\sqrt{s_0^{ab}}$(GeV) & $Z_{ab}$(mb) & $\beta_{T}^{ab}$ & $\beta_V^{ab}$ & $F_{ab}^{(+)}(0)$ \\
\hline
$pp$    & 0.293(26) & 4.64(88) & 34.63(65) & 6.44(35)   & 4.393(41)  & \ 8.1(6) \\
$\pi p$ & 0.293(26) & 5.10(73) & 20.72(39) & 0.143(12)  & 0.0505(12) & 0.06(61)  \\
$Kp$    & 0.293(26) & 5.18(76) & 17.76(43) & 0.408(95)  & 0.687(10)  & \ 2.4(1.0) \\
$pn$   & 0.293(26) & 12.00(75) & 38.90(26) & 2.67(34)   & 3.87(12)  & -15.6(6.8)\\
\hline\hline
\end{tabular}
\label{tabP1}
\end{table}

\begin{figure}[hb]
\begin{center}
\resizebox{0.6\textwidth}{!}{
  \includegraphics{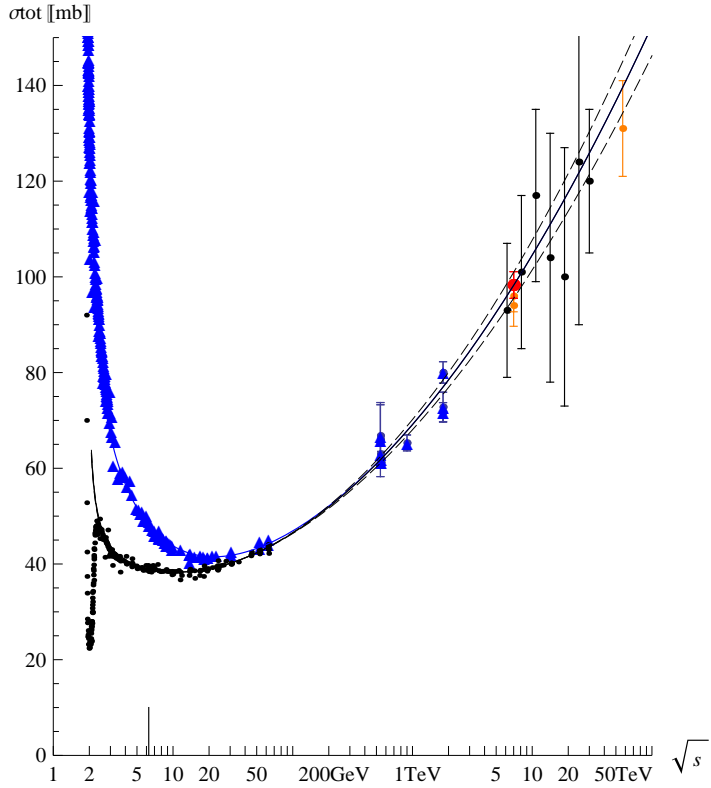}
}
\end{center}
\caption{Fit to the data of $\sigma_{\rm tot}^{\bar pp}$ (blue triangles) and
$\sigma_{\rm tot}^{pp}$ (black circles). The solid lines are our best fit and
the dashed lines correspond to one standard deviation of $B$.
The vertical line on the x-axis represents the lowest energy of the fit region
$\sqrt s\ge 6.27$~GeV corresponding to $k\ge 20$~GeV.
The LHC ATLAS \cite{ATLAS} and CMS \cite{CMS} data (with no extrapolation errors) at 7 TeV (orange) and
the Auger data \cite{Auger} (with only its statistical error) at 57 TeV(orange).
The TOTEM \cite{TOTEM} at 7~TeV is shown by red.}
\label{figP1}
\end{figure}

The number of parameters fit is $6\times 4-6=18$. The fit is very successful
despite the omission of systematic errors of the very high-energy data.
The total $\chi^2$ is $\chi^2/N_{DF}=498.69/(604-18)$, with $\chi^2/N_D$ values of $225.25/245$, $153.95/162$, $63.84/111$ and $55.64/86$ for $\bar p(p)p$, $\pi^\mp p$, $K^\mp p$,
and $\bar p(p)n$ data, respectively.
The results of our best fit to $\sigma_{\rm tot}^{\bar pp,pp}$ are shown in Fig.~\ref{figP1}.
The best-fit values of the parameters are given in Table \ref{tabP1}.
In order to estimate the systematic error of the universal value of $B$,
we shift the central value of $\sigma_{\rm tot}^{pp}$ at 7~TeV by TOTEM \cite{TOTEM}
as $98.3\pm 2.8$~mb.
The corresponding variation of the best-fit value of $B$ is regarded as the systematic error of $B$.
\begin{eqnarray}
B &=& 0.293\pm 0.004_{stat} \pm 0.026_{syst}\ {\rm mb}\ ,
\label{eqP17}
\end{eqnarray}
which is consistent with our previous estimates,
$$B=0.2817(64),~ 0.2792(59)~{\rm mb}~~\cite{BH}~~~ {\rm and}~~~ B=0.280(15)~{\rm mb}~~ \cite{II}.$$
The systematic uncertainty of Eq.~(\ref{eqP17}) is larger than the statistical error
of our previous estimate. We consider this $B$ value is the most conservative estimate
from the present experimental data.

\section{The $\pi\pi$ Total Cross Section}

\subsection{Theoretical Predictions of $\sigma_{\rm tot}^{\pi^\mp \pi^+}$ }

We infer the $\sigma_{\rm tot}^{\pi^\mp\pi^+}(s)$ based on the analyses of
forward $\bar p(p)p$, $\pi^\mp p$, and $K^\mp p$ scattering amplitudes.
Based on the result of the previous section, we can predict $\sigma_{\rm tot}^{\pi^\mp \pi^+}$ at high energy.
By using the relation $s\simeq 2M\nu$ for $ab=ap=pp,Kp,\pi p$ in high-energies, $\sigma_{\rm tot}^{\bar ap,ap}(s)$ of Eq.~(\ref{eqP4}) can be rewritten in the form
\begin{eqnarray}
\sigma_{\rm tot}^{\bar ap,ap}(s) &=& B~{\rm log}^2\frac{s}{s_0}+Z_{ap}+\tilde\beta_{T}^{ap}
         \left( \frac{s}{s_1} \right)^{\alpha_{T}(0)-1} \pm \tilde\beta_{V}^{ap}
         \left( \frac{s}{s_1} \right)^{\alpha_{V}(0)-1}\ \ ,
\label{eqP18}
\end{eqnarray}
where
\begin{eqnarray}
\tilde\beta_{T,V}^{ap} &=& \frac{4\pi\beta_{T,V}^{ap}}{m^2}
        \left( \frac{2Mm}{s_1} \right)^{1-\alpha_{T,V}(0)}\ .
\label{eqP19}
\end{eqnarray}
$s_1$ is introduced as a typical scale for the strong interactions which is taken to be $s_1=1$~GeV$^2$.
It is natural to assume that the universality of $B$ and $s_0$ extend to $\pi\pi$ scattering.
The $\pi^\mp \pi^+$ total cross sections $\sigma_{\rm tot}^{\pi^\mp \pi^+}$ are expected to take the form
\begin{eqnarray}
\sigma_{\rm tot}^{\pi^\mp \pi^+}(s) &=& B~{\rm log}^2\frac{s}{s_0} + Z_{\pi\pi}
+ \tilde\beta_{T}^{\pi\pi}\left( \frac{s}{s_1} \right)^{\alpha_{T}(0)-1}
  \pm \tilde\beta_{V}^{\pi\pi} \left( \frac{s}{s_1} \right)^{\alpha_{V}(0)-1}\ \ ,
\label{eqP20}
\end{eqnarray}
where $B$ and $s_0$ are given by Eq.~(\ref{eqP18}).

The values of $Z_{\pi p},Z_{Kp},Z_{pp}$ in Table \ref{tabP1} approximately satisfy the ratios predicted by the quark model.
\begin{eqnarray}
Z_{\pi p}:Z_{Kp}:Z_{pp}=
20.72:17.76:34.63 \simeq 2:2:3\ .
\label{eqP21}
\end{eqnarray}
By using the quark model meson$/$baryon ratio, $Z_{\pi\pi}$ is
$
Z_{\pi\pi} = \frac{2}{3}Z_{\pi p} = 13.8~{\rm mb}$,
while the $Z_{\pi\pi}$ is also given by
$Z_{\pi\pi} = \frac{Z_{\pi p}}{Z_{pp}} Z_{\pi p} = 12.4~{\rm mb}$,
where the meson$/$baryon ratio is taken to be $Z_{\pi p}/Z_{pp}=0.60$ instead of $2/3$. This assumes that the $Z_{ab}$ terms represent the conventional Pomeron exchange with a unit intercept (and no logarithmic terms) and that its coupling satisfies the Regge factorization.
So our prediction is
\begin{eqnarray}
Z_{\pi\pi} &=&  (12.4\pm 1.4)~{\rm mb}~,
\label{eqP22}
\end{eqnarray}
where the uncertainty is estimated from the difference between the above estimates.
Actually this is the main source of uncertainty for our prediction at very high energy.
Presently  we have no rigorous theoretical way to estimate the accurate value of $s_0^{\pi\pi}$,
hence, we assume for simplicity
\begin{eqnarray}
\sqrt{s_0^{\pi\pi}} &\approx& \sqrt{s_0^{\pi p}}=5.10 \pm 0.73~{\rm GeV}~,
\label{eqPs0}
\end{eqnarray}
where the uncertainty comes from a difference between $s_0^{\pi p}$ and $s_0^{pp}$ for our best fit given in Table~\ref{tabP1}.

The coefficients $\tilde\beta_{T ,V}^{ab}$ take multiplicative forms in terms of Reggeon-$aa(bb)$ couplings
$\gamma_{Raa,Rbb}$ with $\tilde\beta_{T }^{ab}=\gamma_{T aa}\ \gamma_{T bb}$
and $\tilde\beta_{V }^{ab}=\gamma_{V aa}\ \gamma_{V bb}$.
In the case $ab=pp$ and $Kp$, both $f_2(1270)$ and $a_2(1320)$-trajectories contribute via
the tensor-meson term and both $\rho$ and $\omega$-trajectories contribute through the vector-meson term,
while in the case $ab=\pi p$ and $\pi\pi$ only the former trajectories contribute through
the tensor and vector terms. For $ab=pn$, $a_2$ and $\rho$ contributions changes
their signs from $aa=pp$ case.
Using Eq.~(\ref{eqP19}), the values of $\tilde\beta_{T ,V}^{ap}$ are obtained
from Table \ref{tabP1}:
\begin{eqnarray}
\begin{array}{lcllcl}
\tilde\beta_{T }^{\pi p}= &\gamma_{f_2\pi\pi}\gamma_{f_2pp} &=19.4(1.6)~{\rm mb},\ \ \
     & \tilde\beta_{V}^{\pi p}= &\gamma_{\rho\pi\pi}\gamma_{\rho pp} &=6.11(14)~{\rm mb}, \\
\tilde\beta_{T }^{K p}= &\sum_{R=f_2,a_2}\gamma_{RKK}\gamma_{Rpp} &=7.9(1.8)~{\rm mb},\ \ \
     & \tilde\beta_{V}^{K p}= &\sum_{R=\rho,\omega}\gamma_{R KK}\gamma_{R pp} &=13.2(2)~{\rm mb},\\
\tilde\beta_{T }^{p p}= &\gamma_{f_2pp}{}^2+\gamma_{a_2pp}{}^2 &=46.4(2.5)~{\rm mb},\ \ \
     & \tilde\beta_{V}^{p p}= &\gamma_{\rho pp}{}^2+\gamma_{\omega pp}{}^2 &=33.2(3)~{\rm mb}.\\
\tilde\beta_{T }^{p n}= &\gamma_{f_2pp}{}^2-\gamma_{a_2pp}{}^2 &=19.2(2.5)~{\rm mb},\ \ \
     & \tilde\beta_{V}^{p n}= &\gamma_{\rho pp}{}^2-\gamma_{\omega pp}{}^2 &=29.3(9)~{\rm mb}.\\
\end{array}
\label{eqP23}
\end{eqnarray}

The $\gamma$-couplings violates largely the relation of
SU(2) flavor symmetry:
$\gamma_{f_2 pp}=\gamma_{a_2 pp}$, $\gamma_{\rho pp}=\gamma_{\omega pp}$.
Since $\gamma_{\rho\pi\pi}/2=\gamma_{\rho KK}=\gamma_{\omega KK}$,
$\gamma_{f_2\pi\pi}/2=\gamma_{f_2 KK}=\gamma_{a_2 KK}$ from SU(3),
$\tilde\beta_{T ,V}^{\pi p} = \tilde\beta_{T ,V}^{Kp} $
is expected. However, it is also violated in Eq.~(\ref{eqP23}).
On the other hand, for $\pi\pi$ scattering, $\tilde\beta_{T}^{\pi\pi}=\gamma_{f_2\pi\pi}^2$ and
$\tilde\beta_{V}^{\pi\pi} = \gamma_{\rho\pi\pi}^2$. They can be evaluated from
the results of $\bar p(p)p,\bar p(p)n$ and $\pi^\mp p$ of Eq.~(\ref{eqP23}):
\begin{eqnarray}
\tilde\beta_{T}^{\pi\pi} & = & \frac{(19.4\pm 1.6)^2}{(46.4(2.5)+19.2(2.5))/2}{\rm mb}
                     =  11.5\pm 0.9\ {\rm mb},\nonumber \\
\tilde\beta_V^{\pi\pi} &=&   \frac{(6.11\pm 0.14)^2}{(33.2(3)-29.3(9))/2}{\rm mb}
                     =  19\pm 5 {\rm mb} \   ,
\label{eqP24}
\end{eqnarray}
which are compared with the other estimates by using the same method applied to different
inputs; $(\tilde\beta_{T}^{\pi\pi},\tilde\beta_V^{\pi\pi})=(13.39,16.38)$~mb \cite{Sz}
and (8.95(24),21.8(9.0))~mb \cite{Leutwyler}.


\begin{table}[htb]
\caption{Numerical values for the predictions of $\sigma_{\rm tot}^{\pi^\mp\pi^+}$ and their difference
for a range of energies.
Uncertainties from the errors of $\tilde\beta_{T,V}^{\pi\pi}$ decrease with the increasing energies,
and become negligible above $\sqrt s\sim 40$GeV, while the
uncertainties from the errors on $B(=0.293\pm 0.004_{\rm stat}\pm 0.026_{\rm syst}$ mb) and $\sqrt{s_0}(=5.10\pm 0.73$ GeV)
become sizable above this energy. }
\begin{center}
\begin{tabular}{l|ccc}
$\sqrt s$(GeV)  & $\sigma_{\rm tot}^{\pi^-\pi^+}$(mb) & $\sigma_{\rm tot}^{\pi^+\pi^+}$(mb) & difference(mb)\\
\hline
3 & 22.6$\pm 1.4_Z\pm 1.5_{\tilde\beta_{T}}$ & 11.2$\pm 1.4_Z\pm 1.5_{\tilde\beta_{V}}$ &
11.4$\pm 3.0_{\tilde\beta_{V}}$\\
5 & 18.3$\pm 1.4_Z\pm 0.9_{\tilde\beta_{V}}$ & 11.8$\pm 1.4_Z\pm 0.9_{\tilde\beta_{V}}$ &
6.5$\pm 1.8_{\tilde\beta_{V}}$\\
10 & 15.9$\pm 1.4_Z\pm 0.4_{\tilde\beta_{T}}$ & 12.8$\pm 1.4_Z\pm 0.4_{\tilde\beta_{T}}$ &
3.1$\pm 0.8_{\tilde\beta_{V}}$\\
20 & 16.0$\pm 1.4_Z\pm 0.4_{s_0}$ & 14.6$\pm 1.4_Z\pm 0.4_{s_0}$
  & 1.4$\pm 0.4_{\tilde\beta_{V}}$\\
40 & 18.1$\pm 1.4_Z\pm 0.6_{s_0} \pm 0.4_B$ & 17.4$\pm 1.4_Z\pm 0.6_{s_0} \pm 0.4_B$ &
0.7$\pm 0.2_{\tilde\beta_{V}}$\\
50 & 19.1$\pm 1.4_Z\pm 0.7_{s_0} \pm 0.6_B$ & 18.6$\pm 1.4_Z\pm 0.7_{s_0} \pm 0.6_B$ & 0.5\\
100& 23.1$\pm 1.4_Z\pm 0.9_{s_0} \pm 0.9_B$ & 22.8$\pm 1.4_Z\pm 0.9_{s_0} \pm 0.9_B$ & 0.2\\
200 & \multicolumn{2}{c}{28.3$\pm 1.4_Z\pm 1.1_{s_0}\pm 1.4_B$} & 0.0\\
500 &  \multicolumn{2}{c}{37.1$\pm 1.4_Z\pm 1.4_{s_0}\pm 2.2_B$} & 0.0\\
1000 &  \multicolumn{2}{c}{45.1$\pm 1.4_Z\pm 1.6_{s_0}\pm 2.9_B$} & 0.0\\
\hline
\end{tabular}
\end{center}
\label{tabP2}
\end{table}

In summary, $\sigma_{\rm tot}^{\pi^\mp \pi^+}(s)$ are predicted by Eq.~(\ref{eqP20})
with the parameters $B$ from Eq.~(\ref{eqP17}), $Z_{\pi\pi}$ from Eq.~(\ref{eqP22}),
$s_0^{\pi\pi}$ from Eq.~(\ref{eqPs0}), and $\tilde\beta_{T ,V}^{\pi\pi}$ from Eq.~(\ref{eqP24}).
The numerical values of our predictions for several $\sqrt s$-values are given in Table \ref{tabP2}.

\subsection{Comparison with Indirect Experiments}

\begin{figure}[htb]
\begin{center}
\resizebox{0.8\textwidth}{!}{
  \includegraphics{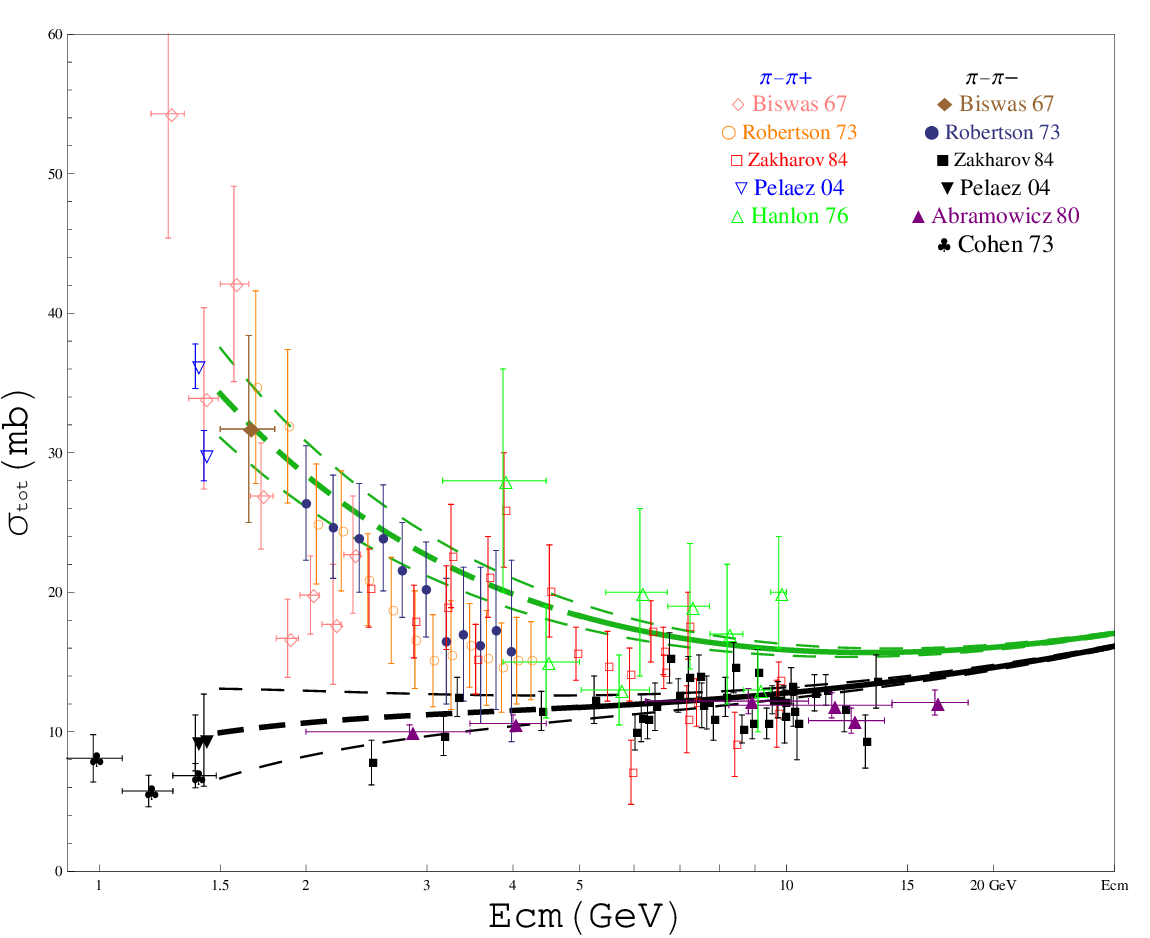}
}
\end{center}
\caption{$\pi^\mp\pi^+$ total cross section (mb) versus $\sqrt s$.
$\sigma_{\rm tot}^{\pi^-\pi^+}$(thick solid green)
and $\sigma_{\rm tot}^{\pi^+\pi^+}$(thick solid black).
Thin dashed lines represent the uncertainty from $\tilde\beta_V^{\pi\pi}$,
which is the largest in the relevant energy region.
}
\label{figP2}
\end{figure}

There are no direct measurements of $\sigma_{\rm tot}^{\pi\pi}$ at present, however, indirect data at low- and intermediate-energy have been extracted
in Robertson73 \cite{Ro}, Biswas67 \cite{Bi}, Cohen73 \cite{Co}, Pelaez03,04 \cite{PY,PY2},
Zakharov84 \cite{Za}, Hanlon76 \cite{Ha}, Abramowicz80 \cite{Ab}.
They are compared with our prediction in Fig.~\ref{figP2}.

The $\chi^2/N_D$ values of our prediction with no free parameters for
the whole data set with $\sqrt s\ge 5$~GeV are not good.
We consider this comes from the quality of the data.
Data are mutually in consistent even in the same collaboration
using different method \cite{Za}.

However,
Our prediction for $\sigma_{\rm tot}^{\pi^-\pi^+}$, by considering the uncertainty from
$\tilde\beta_V^{\pi\pi}$ shown by thin dashed line,  seems to be consistent
with the low-energy regions of Abramowicz80 \cite{Ab} and Zakharov84 \cite{Za}.
As was pointed in Refs. \cite{PY,PY2,Leutwyler},
these data have the natural connection to the low-energy data points by Pelaez03,04 \cite{PY,PY2}
which was recently updated in Ref. \cite{Garcia}.

%
%
%

Our prediction is consistent with the recent other estimates \cite{Sz,Leutwyler,PY2}.
The authors in Ref. \cite{PY} analyze $\sigma_{\rm tot}^{\pi^\mp p}$,
$\sigma_{\rm tot,pp}^{(+)}(=(\sigma_{\rm tot}^{\bar pp}+\sigma_{\rm tot}^{pp})/2)$ and $\pi^\pm\pi^-$ data
simultaneously. All the data, including \cite{Ha,Ab}, as well as the data with very low energies,
$\sqrt s=1.38,\ 1.42$ GeV \cite{PY,PY2}, are included in their fit.
However, at energies of $\sqrt s=1.38,1.42$~GeV,
the Regge theory is not guaranteed to work a priori,
although it seems from Fig.~\ref{figP2}
that it still provides a fairly good description
at those low energies. This suggests that all
sub-leading Regge effects, when combined, result in a rather small contribution.

\section{Discussion and Conclusion}

Our predictions for $\sigma_{\rm tot}^{\pi^\mp\pi^+}$ are shown along with the results of our best fit for $\sigma_{\rm tot}^{\pi^\mp p}$ and $\sigma_{\rm tot}^{\bar p(p)p}$ in Fig.~\ref{figP4}.
The difference in normalization of these curves is determined by the $Z_{ab}$, $Z_{pp}>Z_{\pi p}>Z_{\pi\pi}$,
while their increase with energy is described by the universal value of $B$, Eq.~(\ref{eqP17}).

There are a few comments as our concluding remarks:

\begin{itemize}
\item We have previously predicted the $\sigma_{\rm tot}^{pp}$ for LHC and cosmic-ray energies in \cite{BH,II}.
Now our previous predictions can be tested by using the new experimental data \cite{ATLAS,CMS,Auger,TOTEM} shown
in Table~\ref{tabP3}. The result of the fit in the present work is also shown,
together with the predictions at $\sqrt s=8,14$~TeV.
Our predictions are in good agreement with the experiments.

Our result is also compared with the other predictions \cite{COMPETE,Aichili,PY}
at $\sqrt s=$14~TeV in Table~\ref{tabP4}. All the models give consistent results within
their uncertainties.
The central value of B in the present analysis becomes somewhat larger
than our previous estimate as can be seen in Eq.~(\ref{eqP17}).
This larger $B$ value comes from the TOTEM measurement \cite{TOTEM}, of which value
includes a large systematic uncertainty.
As a result our present prediction at $\sqrt s=$14~TeV becomes almost the same as
that of COMPETE collaboration as can be seen by Table~\ref{tabP4}.
Our previous prediction based on duality
constraint will be tested more strictly in the future LHC experiment $\sqrt s=8$~TeV.

\begin{table}[htb]
\caption{Comparison of our previous predictions of $\sigma_{\rm tot}^{pp}$(mb) with the new experiments by
ATLAS \cite{ATLAS}, CMS \cite{CMS}, TOTEM \cite{TOTEM} and Auger \cite{Auger}.
The result of the fit in the present work is also given.
For the derivation of the experimental values, see the text.
The numbers with parentheses are the fit results, and the others are predictions. }
\begin{center}
\begin{tabular}{c|lll|l}
$\sqrt s$(TeV) & $\sigma_{\rm tot}^{pp}$ (BH) \cite{BH} &  $\sigma_{\rm tot}^{pp}$ (II) \cite{II} & this work &  $\sigma_{\rm tot}^{pp}$(mb) (exp.)\\
\hline
7  &  95.4$\pm$1.1 &  96.0$\pm$1.4 & (98.2$\pm$2.7) & 96.0$\pm$3.3$_{exp.}$$\pm$9.5$_{extr.}$ : ATLAS  \\
   &               &       &  & 94.0$\pm$2.8$_{Syst.}$$\pm$3.3$_{Lum.}$$\pm$5.5$_{Extr.}$ : CMS  \\
      &               &               &  & 98.3$\pm 0.2_{stat}\pm 2.8_{syst}$ : TOTEM  \\
8  &  97.6$\pm$1.1 &  98.2$\pm$1.5 & 100.6$\pm$2.9 & \\
10 & 101.4$\pm$1.2 & 102.0$\pm$1.7 & 104.5$\pm$3.1 & \\
14 & 107.3$\pm$1.2 & 108.0$\pm$1.9 & 110.8$\pm$3.5 & \\
57 & 134.8$\pm$1.5
  & 135.5$\pm$3.1 & (139.6$\pm$5.4) & 131$\pm$10$_{Stat.}\stackrel{+12}{\scriptstyle -16}_{Syst.}$$\pm$2$_{Glauber}$ : Auger\\
\hline
\end{tabular}
\end{center}
\label{tabP3}
\end{table}

\begin{table}[htb]
\caption{Comparison of our prediction of $\sigma_{\rm tot}^{pp}$(mb) at $\sqrt s=$14~TeV
with the other works: COMPETE \cite{COMPETE}, AGGPSS \cite{Aichili}, and PY \cite{PY}.
$*$ This uncertainty is quoted form the largest and smallest values given in Table I of Ref. \cite{Aichili}. }
\begin{center}
\begin{tabular}{c|llll}
$\sqrt s$(TeV) & this work & COMPETE \cite{COMPETE} &  AGGPSS \cite{Aichili} & PY \cite{PY}\\
\hline
%
14 & 110.8$\pm$3.5 & 111.5$\pm 1.2\stackrel{+4.1}{\scriptstyle -2.1}$
     &  100.3$\stackrel{+10.2}{\scriptstyle -12.5}^*$  & 104$\pm$4, 113$\pm$4 \\
\hline
\end{tabular}
\end{center}
\label{tabP4}
\end{table}

\begin{figure}[h]
\begin{center}
\resizebox{0.6\textwidth}{!}{
  \includegraphics{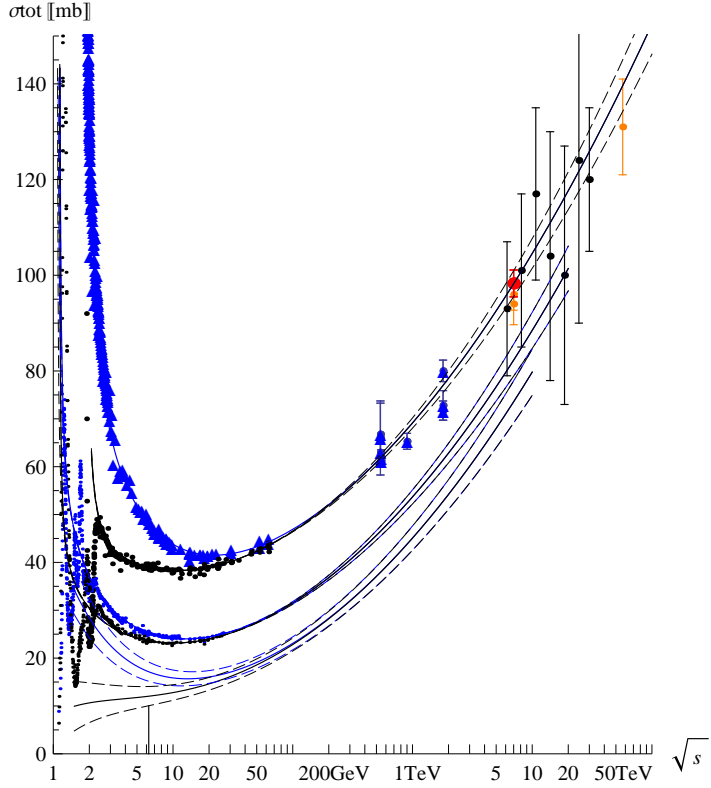}
}
\end{center}
\caption{Total cross sections(mb) versus $\sqrt s$.
Solid lines are $\bar pp,pp$, $\pi^-p,\pi^+p$, $\pi^-\pi^+, \pi^+\pi^+$ from up-to-down.
$\bar pp,pp$ and $\pi^-p,\pi^+p$ are our best fit, while $\pi^-\pi^+,\pi^+\pi^+$ are our predictions.
The dot-dashed lines represent the upper(lower)-limit of our predictions of
$\sigma_{\rm tot}^{\pi^-\pi^+}$($\sigma_{\rm tot}^{\pi^+\pi^+}$).
Dashed lines for $\bar p(p)p$ and $\pi^\mp p$ represent the uncertainties of our predictions
which are obtained from the errors of $B$ and $Z_{\pi\pi}$.
}
\label{figP4}
\end{figure}

\item The COMPETE collaboration assumed the universality of $s_0$ in their fit \cite{COMPETE,PDG}.
We have tested this $s_0$ universality \cite{II}.
By applying the further constraints $s_0^{pp}=s_0^{\pi p}=s_0^{Kp}$ in our analysis
of $\bar p(p)p,\pi^\mp p,K^\mp p$ data(not including $\bar p(p)n$ data),
we obtain $B=0.299(8)$ mb and the universal $\sqrt{s_0}=5.59(30)$ GeV,
which are consistent with $B=0.308(10)$ mb and $\sqrt{s_0}=5.38(50)$ GeV of COMPETE collaboration \cite{COMPETE,PDG}.
However, the $\chi^2$ value of this additional constraint on $s_0$,
$\chi^2/N_{DF}=438.53/(517-11)$, becomes worse by 7 units (for extra 2 constraints)
compared with our best fit $\chi^2/N_{DF}=431.48/(517-13)$ with no constraint on $s_0$.
The data seem to favor the fit without $s_0$-universality although the $\chi^2$ improvement
is not remarkable in this case.

\item
$s_0^{pp}=s_0^{pn}$ is further assumed in the COMPETE analysis.
In mini-jet model\cite{jet}, the $c_2$ coefficient
is described by gluon-gluon scattering, and thus it is the same as $pp$ and $pn$ system, while the
$c_1$ coefficient includes the effect of quark-gluon scattering, thus, it is generically
different between $pp$ and $pn$. Correspondingly, $B_{pp}=B_{pn}$ consistent with the universality,
while $s_0^{pp}\neq s_0^{pn}$. Our best fit value $\sqrt s_0^{pn}=12.0$~GeV is slightly larger
than $\sqrt s_0^{pp}=4.6$~GeV. If we take $s_0^{pn}=s_0^{pp}$ as one more constraint,
the resulting $\chi^2$ for the $(\bar p)pn$ data is $\chi^2/N_{DF}=69.3/(86-4)$
which is 14 units larger than the
original value $\chi^2(pn\ {\rm data})$, $\chi^2/N_{DF}=55.6/(86-5)$.
The reduced $\chi^2$ is less than unity also for the $s_0$-universal fit, but
the experimental data prefer the non $s_0$-universal fit.
So we did not adopt the $s_0$-universality in the present analysis.

\end{itemize}

\section*{Acknowledgments}

F.H. is supported by the National Science Foundation award 096906.
M.I. is very grateful to Dr. R. Kaminski for informing the $\pi\pi$ data and giving  crucial comments.
This work is supported in part by KAKENHI (2274015, Grant-in-Aid for Young Scientists(B)) and in part by grant
as Special Researcher of Meisei University.
C.S.K.  is supported  by the NRF grant funded by  Korea government of MEST
(No. 2011-0027275), (No. 2011-0017430) and (No. 2011-0020333).
\\




%
%
%

\end{document}